# Discovering Business Rules from Business Process Models

Thanh Thoa Pham Thi, Markus Helfert, Fakir Hossain, Thang Le Dinh

Abstract: Discovering business rules from business process models are of advantage to ensure the compliance of business processes with business rules. Furthermore it provides the agility of business processes in case of business rules evolution. Current approaches are limited on types of rules that can be discovered. This paper analyses the expression power of some popular business process modelling languages in embedding business rules in its presentation and provides indicators to extract various types of business rules from business process models.

INTRODUCTION

Business processes can be considered as the centre of the business of an organisation. They are related to different aspects of a business such as business structure, business activities, events, organizational units and interactions between those aspects to achieve goals defined by the organisation [6].

Business rules (BR) are the integral part of the business which represents business policies, internal regulations, external regulations, procedures or standards set up by the professional corps which the organization must comply with. A BR is "a statement that defines or constrains some aspects of a business" (Business Rules Group). Alternatively BR is "a proposition about business things, relationships between them and operations applied to them, from the business enterprise viewpoint" [2]. In other words, BRs control the business; consequently, BP depends directly on BR [11].

Business requirement analysis for information system development demands to capture different aspects of business by means of modelling languages. Actually it is convinced that there is still a lack of documentations on BR, and if BRs are not acquired systematically and completely, they cannot reflect real condition of business environment. Consequently, developed applications may not meet business requirements [3]. This issue clearly relates to the incompliance of BP models which results from business requirement analysis with BR.

Let's take an example of an online hotel booking service provided by a travel agency to illustrate the relation of business rules and business processes: "The agency collaborates with many hotels and provides the hotel booking service to clients. A client makes a hotel booking through the agency website and some other services provided by hotels. The client must provide his credit card information to the agency (via its website) when he books. Once the client confirms the booking, the agency automatically sends a confirmation email to the client as well as to the corresponding hotel. The cancellation policies may be not the same for every hotel, but in general it is stated that cancellations before 24 hours (or 48 hours) to check-in date and time (i.e. earliest check-in time) are free of charge; otherwise the first night's fee will be charged. Once the booking is cancelled, an email needs to be sent to the client and another one needs to be sent to the hotel to confirm the cancellation".

The BP model of this service must describe the action Charge the first night fee following a late cancellation event. This business logic is an implementation of the business rule: On receiving the cancellation request If clients make the late cancellation Then their first night fee is charged.

Current approaches for discovering business rules from business processes are limited on type of rules that can be discovered. At the same time, a second stream of research aims to integrate business rules into business process [13][15]. In [14], Morgan explains general indicators to find business rules in business environment, for example business rules can find in "features defined by external agencies", "automated business decision making", "systematic variations among organizational units", "entities with multiple states", "derivations, or calculations", etc. However, there is not a systematic guidance to discover business rules.

In this paper, we propose some indicators for discovering BR from BP by analyzing how and which BR are incorporated in BP. It makes two important contributions:
- Firstly, it allows to verify the compliance of BP with BR at the conceptual level,
- Secondly, it allows to coordinate BP and BR effectively in case of evolving BR and/or BP.

The remainder of the paper is organized as follow: in the next section, we resume basic concepts of business rules and business process models. Later we discuss some related work on discovering business rules and some work on integrating business rules into business processes. Following that section is the presentation of indicators to extract business rules with illustration of the online hotel booking example. Next part deals with evaluation of the expression power of Petri Nets and EPC business process modelling languages based on specifications presented in the previous section. The paper is finished with its conclusions and some discussions.

BUSINESS RULES AND BUSINESS PROCESS

Business Rules Group has classified business rules into three categories: Structural assertions, action assertions and derivations [8]. Structural assertions concern on business structures such as data object, organization structure, and resource. Action assertions concern rules on triggering business activities, business behaviour, and process flow. Derivations concerns rules on inferences, calculations [8].

BR is normally described by natural language, which can be reconstructed as Event-Condition -Action form [4][9] which describes statements On <Event> If <Condition is satisfied> Then Do <Action>. Business rules can be also in form of <Subject> Must <constraints> [14], or in form of If- Then expression.

BP captures business activities to achieve an explicit goal of an organization and their relationship with the resources participating in the process requirements [12]. There are many languages developed for BP modelling. In this section we present very briefly two popular languages: Petri Net (PN) based for workflow modelling [19], which are implemented in WFMS as COSA, INCOME, and Event-Driven Process Chain (EPC) modelling technique [18] implemented in SAP R/3, ARIS, Microsoft Visio.

PN based techniques for workflow (WF) modelling normally separate the process dimension from the structure of organization and the resource within organization (by whom) to reduce the complexity [18]. Process modelling with Petri-Nets describes which tasks are performed in what order with the following constructs: State, transition, token and condition. State represents conditions of WF. Transitions represent tasks. A token represents the WF state of a single case, or business process instance. A condition may represent an event, or a data status which allows to enable a task, but it is not explicitly specified. There is not explicitly data description in the model. Routing constructs include sequential routing, AND-split, AND-join, XOR-split/join. The triggering of a task corresponds to events which may be external conditions such as a message or on a specific time.

The Event Driven Process Chain (EPC) method is also based on the concepts of stochastic networks and PNs [18]. EPC is able to incorporate functions, data, organizational structure, resource in the model. The main constructs of EPC which are related to our work as follows. Function corresponds to a task or a process step in the business process. Event represents conditions, circumstances or situation before or after a function is performed. Information, material or resource object describes business objects which may be input/output data of function. Logical connectors specify the control the flow which connects functions and events.

RELATED WORK

In literature, there are some approaches exploring the correlation between business process and business rules.  The work in [7] proposes a method for discovering business rules through the use of business process mining and data mining based on event logs. In contrast to the focus of this paper, this approach is interested in two kinds of rules: event-condition-action and authorization action.

Meanwhile the approach presented in [16] aims to enable agile business process, it proposed patterns of business rules can be extracted from business process and using business rules to control business process. This approach deeply explores business rules extracted around the decision node (XOR connector) in a business process model.,

In contrast, the work in [5][17] use action assertions BR to develop the workflow model. The approach proposed by [13] tries to integrate more business rules into business processes by introduce a domain constraint repository which is able to verify the semantic correctness of business process after evolution. The constraints they are focusing are the exclusive and dependency constraints of tasks.  The domain constraint repository captures the semantic of domain to indicate when and what tasks are exclusive or dependent to the other one. In [15], some extended constructs of BPMN are introduced such as Operating Condition and Control case so that non functional requirement of business processes including business constraints can be able to describe.  Some other approaches use business rules to control over business process deployment [1]. These approaches are not concerned to discover business rules from business processes. Different to [7], our approach is based on the business process models documentation to extract business rules. By this way, more types of rules can be discovered compared with the work in [7] and [10].

INDICATORS TO EXTRACT BUSINESS RULES

By analyzing BP models described with different BP modelling techniques and by observation, we identify the following patterns where business rules are likely incorporated in business processes:

• Pattern 1: Decision point, automate decision. The incorporated business rules tell basing on what condition the decision is made and it defines the condition

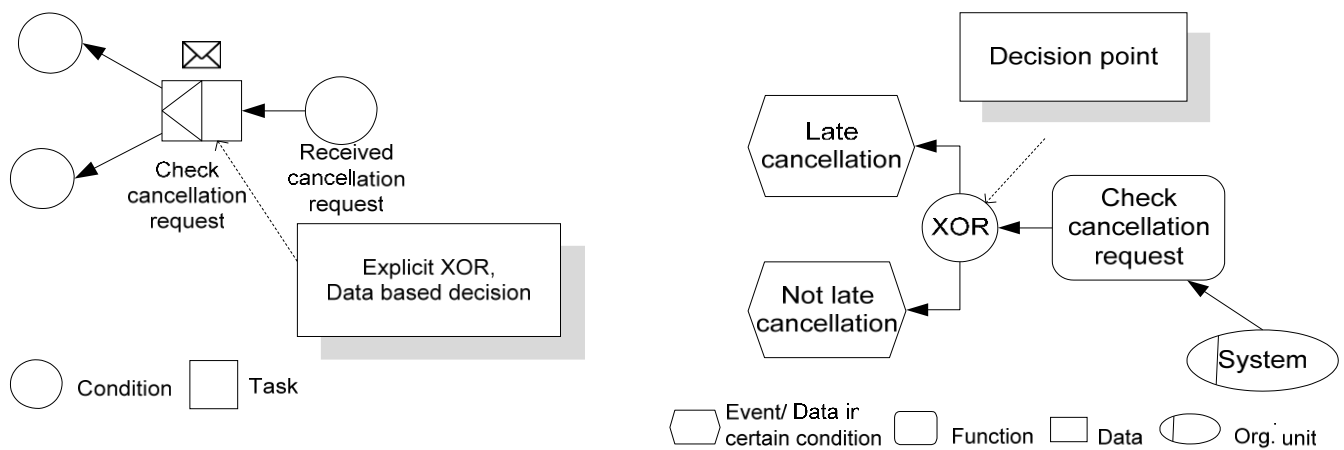

Figure1. Pattern 1- Petri Nets Notation (a)          EPC Notation(b)

A decision point corresponds to the XOR connector logic. A manual decision point is often made based on subjective judgement. A programmed decision can be automatically made based on rules indicating how the decision is made. The rules are applied on data value involved in the activity. This case has two sub-cases following the programmed decision which corresponds to the activity: (1.1) a decision on what data to be outputted, or (1.2) a decision on what action to be proceeded.

- In case (i), we have the following derivation assertion(1.1): If<Condition C1 > Then <Data 1 is produced>, If <Condition C2> Then <Data 2 is produced>
- In case (ii), we have the following action assertions(1.2):
  o On <Event> If <Condition 1> Then Do<Activity 1>
       If <Condition 2> Then Do <Activity 2>

Let's see figure 1, analyzing Check Cancellation request activity, we have the following rules:
- If <Received Cancel request date-time is less than 24 hours to check-in date> Then <Late cancellation is produced>
- If <Received Cancel request date-time is more than 24 hours to check-in date> Then <Not Late cancellation is produced>

• Pattern 2: Logic connectors. Logic connectors other than XOR can present a rule on an action or a rule between data related to that action

There are possible these types of rule. (2.1): [On<Event>] If <Condition> Then <Action>, (2.2): If <data1 is in a condition 1> then <data2 is in a condition 2>

Condition concerns a logic expression on input/output data value of the activity. Action corresponds to triggering the activity.

Figure 2 presents a part of BP model of online hotel booking example. The connector logic AND before two activities Send confirmation email to client and Send confirmation email to the hotel.

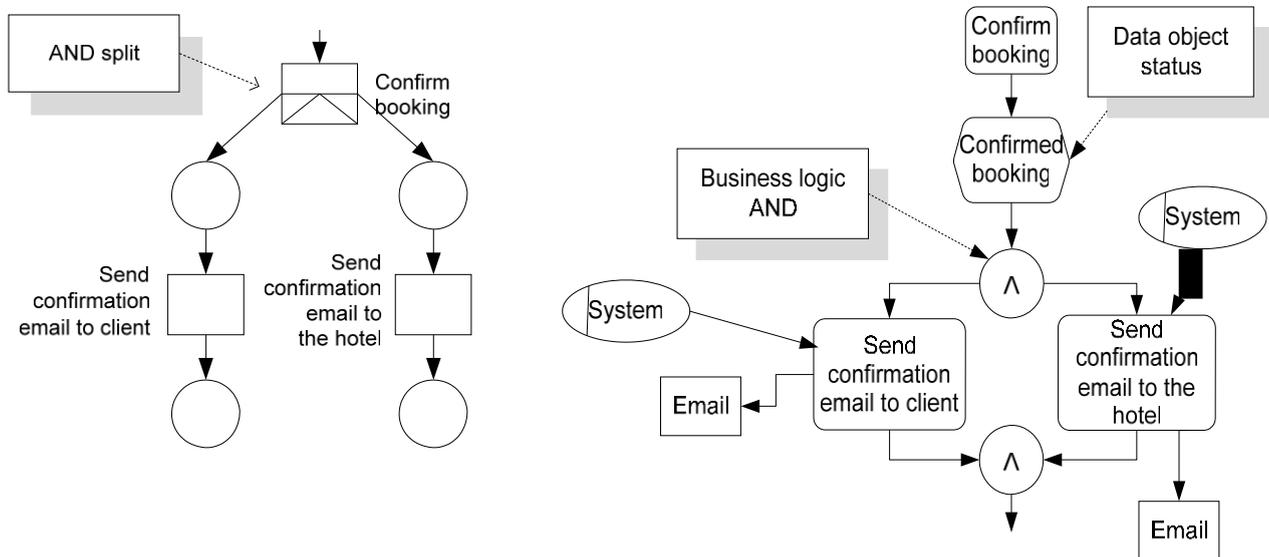

Figure 2. Pattern 2 – PN notation (a)                      EPC Notation (b)

This rule (type 2.1) can be captured: If <Booking is confirmed> then <Send confirmation email to the client> AND <Send email to confirm cancellation to the hotel> (the condition Booing is confirmed is also an event), and a rule type 2.2: If <a booking is in confirmed status> then <there exist two confirmation emails related to that booking>.

•Pattern 3: Data objects within states

One of ways to monitor the business logic of business processes is based on data produced by those business activities. The order of data produced represented the order of business activities. Therefore it is necessary to capture the order of state change of data objects and constrains this order as a rule, that allows to ensure the consistency of data and process [16]. The rule pattern is (3.1): If <data1 is in a status x> then <it must already spent the status y>, or in assertion form (3.2) <object cannot obtain a status y from a status x> (in case y is a previous state of x). This rule can be interpreted in several ways depending on how an object status is defined. We explored this type of rule in [16].

In this example, a complete specification of the business process will show a booking confirmation (Figure 2) is ahead of a booking cancellation (Figure 1). The corresponding rule is captured: " A booking can change from confirmed status to cancelled status but not inverse".

• Pattern 4: Organizational roles implies authorization rules

There are generally business rules on authorization that indicate who is responsible for activity performing. This kind of rule is described in the form (4.1) <Subject> Must <Constraint>.

Basing on the specification in Figure 2(b), this rule is obtained: "the system must send an email to the client " (in term of responsibility).

EXPRESSION POWER OF BUSINESS PROCESS MODELLING LANGUAGES IN BUSINESS RULE PRESENTATION

Investigating the expression power of business process modelling language in business rule presentation helps to evaluate how much and which business rules can be incorporated in business process.

Based on the modelling of the example above with Petri Nets and EPC, we discuss how business rules are incorporated in the two BP models as following.

The WF modelling based on Petri-Nets focuses on the logic of tasks, however not on input and output data of tasks. Therefore action assertions are described clearly in the specification, but structural assertions on data object are not explicit included. Petri Nets based method focuses on tasks description and transition from one task to others ones (task flow control), but data involved in each task is not really described.  A transition is a condition which may be data or event of which their semantic is not clearly defined. Consequently, rules concerning data object state are not clearly incorporated. Meanwhile, that is not the case in EPC. The EPC allows to explicitly describe business logic and data involved in functions. Indeed, EPC allows to model organizational units assigned to each function, therefore it is able to describe authorization rules In other words, EPC technique allows to capture more rules because it describes more information in the model. The following table summarizes the expression power of these two models on describing business rules.

Table 1. Expression power of BP modelling technique on incorporated business rules

| Rules templates | Petri Nets based | EPC |
|---|---|---|
| Rules concerning programmed decision (XOR connector) | ● | ● |
| Rules concerning other connector logics | ● | ● |
| Rules concerning data object state | ○ | ● |
| Authorization rules | ○ | ● |

Notation:  Filled circle: covered rules, Empty circle: not covered rules

5. CONCLUSION

Business rules are often incorporated in business process models to represent the business logic and to control business activities according to organization policies or rules. It depends on the expression power of each business process modelling language that how much business rules can be described in a business process model. Discovering business rules from business processes helps to verify if the rules incorporated in business process are correct. In other words it allows to verify the compliance of the business process and organization rules. Furthermore it allows effectively to manage business rules and business process in case of evolving business and rules. The related approaches are limited at the types of discovered rules. In this paper, we have presented our approach for discovering more types of business rules from business processes. By analyzing two selected BP modelling techniques, we developed patterns where business rules are likely involved. These patterns can act as rule templates for business rules specification. Among these patterns we have developed a systematic technique and tool for discovering rules of pattern 3, for other patterns there needs to be more investigation. Our work shows that there is a variation of rules embedded in business process presentation depending on business process modelling languages. This issue should be considered in any approach of (semi-) automatic discovering business rules from a business process specification, which is also our further research in the future.

REFERENCES

[1] Ali, S., T. Torabi, B. Soh, Rule component specification for business process deployment, in proceedings of 18th International Workshop on Database and Expert Systems Applications, 2007

[2] ANSI, Business Rules for the Enterprise Viewpoint of RM-ODP. ANSI X3H7-96-07R2, 7 December, 1996

[3] Bajec, M., M. Krisper, R. Rupnik, Using business rules technologies to bridge the gap between business and business applications, in: G. Rechnu (Ed.), Proceedings of the IFIP 16th World Computer Congress 2000 Information Technology for Business Management, Beijing, China, 2000, pp. 77–85

[4] Bubenko, J., Jr, & Wangler, B. Objectives driven capture of business rules and of information systems requirements. Proceedings of the International Conference on Systems, Man and Cybernetics, 1993, 670–677

[5] Endl, R., G. Knolmayer, M. Pfahrer, Modeling Processes and Workflows by Business Rules, 1st European workshop on Workflow and Process Management WPM'98, 1998

[6] Ericksson, H., M. Penker, Business Modeling with UML, Business Patterns at Work, Wiley Computer Publishing, 2000

[7] Halpin, T., J. Krogstie, S. Nurcan, E. Proper, R. Schmidt, P. Soffer and R. Ukor, Discovering Business Rules through Process Mining, Enterprise, Business-Process and Information Systems Modeling, 10th International Workshop, BPMDS, 2009.

[8] Hay, D., K.A. Healy, Defining Business Rules_What Are They Really? Technical Report Rev 1.3, the Business Rules Group, July 2000

[9] Herbst, H., A meta-model for specifying business rules in system analysis. Proceedings of CaiSE'95, 1995,186–199.

[10] Graml, T., R. Bracht, M. Spies, Patterns of Business Rules to Enable Agile Business Process, 11th IEEE International Enterprise Distributed Object Computing Conference, 2007.

[11] Kadir, W., W.M.Norman and P. Loucopoulos, Relating Evolving Business Rules to Software Design, Journal of systems architecture, 2003.

[12] Lankhorst, M., et al., Enterprise Architecture at Work, Modelling, Communication and Analysis, Springer 2009

[13] Ly, L. T., S. Rinderle and P. Dadam, Integration and verification of semantic constraints in adaptive process management systems, Journal of Data & Knowledge Engineering, Vol. 64, 2008, pp.3-23

[14] Morgan,T., Business Rules and Information Systems: Aligning IT with Business Goals, Addison-Wesley, Boston, MA, 2002.

[15] Pavlovski, C. J., J. Zou, Non functional requirement in business process modeling, 2008

[16] PhamThi, T.T., M. Helfert, Discovering Dynamic Integrity Rules with a Rules-Based Tool for Data Quality Analyzing, ACM CompSysTech'10, Sofia, June 2010

[17] Rowe, A., S. Stephens, Y. Guo, The use of business rules with workflow systems, in: W3C Workshop on Rule Languages for Interoperability, 2005

[18] Scheer, A. W., ARIS – Business Process Modeling, the Third Edition, Springer, 2000

[19] Van der Aalst, W.M.P.,The Application of Petri Nets to Workflow Management. The Journal of Circuits, Systems and Computers, 1998, 8(1):21–66.